\documentclass[iop]{emulateapj}
\usepackage{courier,graphicx,epsfig,color,colordvi,natbib,epstopdf}
\usepackage[normalem]{ulem}
\usepackage[colorlinks=true,urlcolor=blue,citecolor=blue,linkcolor=blue]{hyperref}
\usepackage[flushleft]{threeparttable}

\begin{document}

\title{Height dependence of the penumbral fine-scale structure in the inner solar atmosphere}

\author{Mariarita Murabito \altaffilmark{1}, I. Ermolli\altaffilmark{1}, F. Giorgi\altaffilmark{1}, M. Stangalini\altaffilmark{1}, S.L. Guglielmino\altaffilmark{2}, S. Jafarzadeh\altaffilmark{3,4}, H. Socas-Navarro\altaffilmark{5,6}, P. Romano\altaffilmark{7}, F. Zuccarello\altaffilmark{2}}\email{mariarita.murabito@inaf.it}

\altaffiltext{1}{INAF - Osservatorio Astronomico di Roma, Via Frascati,33, 00078 Monte Porzio Catone, Italy}
\altaffiltext{2}{Dipartimento di Fisica e Astronomia - Sezione Astrofisica, Universit\`{a} degli Studi di Catania, Via S. Sofia 78, 95123 Catania, Italy}
\altaffiltext{3}{Rosseland Centre for Solar Physics, University of Oslo, P.O. Box 1029 Blindern, NO-0315 Oslo, Norway}
\altaffiltext{4}{Institute of Theoretical Astrophysics, University of Oslo, P.O. Box 1029 Blindern, NO-0315 Oslo, Norway}
\altaffiltext{5}{Instituto de Astrof\'isica de Canarias, 38205, C/ Via L\'actea s/n, La Laguna, Tenerife, Spain}
\altaffiltext{6}{Departimento de Astrofisica, Universidad de La Laguna, 38205 La Laguna,Tenerife, Spain}
\altaffiltext{7}{INAF - Osservatorio Astrofisico di Catania, Via S. Sofia 78, 95123 Catania, Italy.}

\shorttitle{}
\shortauthors{Murabito et al.}

\begin{abstract}

We studied the physical parameters of the penumbra in a large and fully-developed sunspot, one of the largest over the last two solar cycles, by using full-Stokes measurements taken at the photospheric \ion{Fe}{1} 617.3 nm and chromospheric \ion{Ca}{2} 854.2 nm lines with the Interferometric Bidimensional Spectrometer. Inverting measurements with the NICOLE code, we obtained the three-dimensional structure of the magnetic field in the penumbra from the bottom of the photosphere up to the middle chromosphere. 
We analyzed the azimuthal and  vertical gradient of the magnetic field strength and inclination. Our results provide new insights  on the properties of the penumbral magnetic fields in the chromosphere at atmospheric heights unexplored in previous studies.  We found signatures of the small-scale spine and intra-spine structure of both the magnetic field strength and inclination at all investigated atmospheric heights. In particular, we report typical peak-to-peak variations of the field strength and inclination of $\approx$ 300 G and $\approx 20^{\circ}$, respectively, in the photosphere, and of $\approx$ 200 G and $\approx 10^{\circ}$ in the chromosphere. Besides, we estimated \textbf{the} vertical gradient of the magnetic field strength in the studied penumbra: we find a value of $\approx$ 0.3 G km$^{-1}$ between the photosphere and the middle chromosphere. Interestingly, the photospheric magnetic field gradient changes sign from negative in the inner to positive in the outer penumbra. 

\end{abstract}


\section{Introduction}{
\label{sec:introduction}

Sunspots, which are the most prominent feature of the solar photosphere, are primary manifestations of the Sun's magnetism  
\citep[][]{rempelschliche2011lrsp,Lidiareview15}.  

Penumbrae are integral parts of sunspots \citep[][]{solanki2003,borreroichimoto2011,Tiwari2017}. Since earlier telescopic observations that showed the complex filamentary nature of these, penumbrae have been the focus of numerous investigations. 
In recent years, observations taken with state-of-the-art space- and ground-based telescopes have been employed to study the formation of penumbrae by, e.g.,  \citet{Schl10b,Schl12,Rom13,Rom14,Murabito16,Murabito17,Murabito18}  and to investigate their three-dimensional (3D) magnetic structure by, e.g., \citet[][]{Tiwari2013,Joshi2016,Joshi2017a,Tiwari2017}. 

In the photosphere, penumbrae consist of nearly radially aligned filaments where strong and weak magnetic fields  are interlaced with each other along the azimuthal direction.  Indifferently from their position, penumbral filaments exhibit bright heads nearest to the umbral region, and dark cores along their central axes  \citep[]{Tiwari2013}. The filaments' heads  show fields with same polarity as the  umbrae, which are  enhanced ($\approx$1.5-2 kG) and more vertical ($\approx$35$^{\circ}$) than  the fields observed along the axes of filaments. The latter fields are comparatively weaker ($\approx$1 kG) and more horizontal ($\approx$70$^{\circ}$) also with respect to the fields in filaments' tails, which are rather strong ($\approx$2-3.5 kG) and vertical fields of opposite polarity than umbrae \citep[]{Tiwari2013}. This field arrangement, which is usually described in terms of interlaced spines (more vertical and stronger fields) and intra-spines (more horizontal and weaker field), 
 is typical of all of the photospheric heights, but more prominent in the lower layers. 
The same field arrangement supports some models presented in the literature, e.g. the  ``uncombed'' and ``interlocking'' penumbra models proposed by \citet[][]{SolankiMontavon1993} and \citet[][]{thomas2002}, respectively.
According to  these models, the magnetic field strength in penumbrae decreases with atmospheric height, as well as its inclination; i.e., the field becomes more vertical when moving to higher atmospheric heights.

Recently, \citet{Balthasar2018} presented an extensive review about the unsolved problems of the estimation of the magnetic field gradient in sunspots, exploring the different techniques employed and the reasons for the discrepancy of the results presented in the literature. 
For example, \citet{Ruedi1995} found that the vertical gradient of the magnetic field decreases outwards in the sunspot, with values of  0.1-0.3 G km$^{-1}$ in the outer penumbra, and with the height in the atmosphere.   
Other authors, specifically \citet{Westendorp2001}, \citet{borreroichimoto2011}, and \citet{orozco2015},  reported signatures of  magnetic fields forming canopy-like structures in the middle and outer penumbrae, while \citet{Mathew2003}, \cite{sanchezcuberes2005}, and \citet{balthasargomory2008} reported that the field simply increases with depth everywhere in sunspots.  \citet[][]{Tiwari2013,Tiwari2015} did not find any evidence for the canopy-like structure, but reported of a reversed magnetic field gradient in the inner penumbra.
\citet{Joshi2016} found that the magnetic field of the penumbra is more vertical in the upper chromosphere compared to that in the photosphere. Moreover, they found that the inclination of the field varies along the azimuthal direction in both the photosphere and upper chromosphere, while they reported  small-scale spine/intra-spine fluctuations of the field strength at photospheric heights only. Furthermore, 
 \citet{Joshi2017a} 
found that the vertical gradient of the field strength displays large spatial fluctuations in the photosphere, changing its sign even on small scales.  In particular,  \citet{Joshi2017a} found that 
the vertical gradient of the field 
 is always positive in spines, while in the part of the filament closer to the umbra it is positive and surrounded by negative gradient at the sides of the filament in the lower atmospheric layers, being always negative in the upper layers. \citet{Joshi2017a} also reported that the tail of filaments has a positive field gradient both in the lower and upper photospheric layers. 



It is worth nothing that the above current understanding on the small-scale 3D magnetic structure of penumbral filaments derives from analysis of data taken through several photospheric \ion{Fe}{1}, \ion{Si}{1}, and \ion{Ca}{1} lines\footnote{Specifically, the \ion{Fe}{1} 630.1, \ion{Fe}{1} 630.2 nm, \ion{Fe}{1} 1564.8 nm, \ion{Fe}{1} 1565.3 nm, \ion{Fe}{1} 1078.3 nm, \ion{Si}{1} 1078.4 nm , \ion{Si}{1} 1078.7 nm, \ion{Si}{1} 1082.7 nm, and \ion{Ca}{1} 1083.3 nm lines.}, 
but one  chromospheric diagnostic only, the \ion{He}{1} triplet at 1083.0 nm.   
The latter spectral region, which offers a unique tool to study photospheric and chromospheric fields simultaneously, is strongly influenced by EUV coronal irradiation penetrating the atmosphere deep into the upper chromosphere,  to a height that hydrostatic atmosphere models indicate to be around 2000 km above the photosphere \citep{Avrett1994}, where the He I  1083.0 nm spectrum is thought to be formed. From analysis of penumbral observations taken in the spectral region including the He~I triplet, \citet{Joshi2017b} reported the presence of the small-scale spine/intra-spine structure of the magnetic field strength and inclination at photospheric heights between log$\tau$=0 and log$\tau$=-2.3, where  log$\tau$ is the logarithm to the power of 10 of the optical depth, and of the same but attenuated structure only for the field inclination in the chromosphere sampled with the \ion{He}{1} triplet. However, due to the current limited observations and modelling capabilities for investigating the chromospheric heights, the magnetic field structure of penumbrae above the photosphere remains poorly understood.

Recently, \citet[][]{Joshi2018} studied the variations of magnetic field in a sunspot from the photosphere to the chromosphere by using a different chromospheric diagnostic than in the  previous works, specifically observations taken in the Ca II 854.2 nm line. According to \citet[][]{Quintero}, the \ion{Ca}{2} 854.2 nm line is mostly sensitive to  the atmospheric layers enclosed in the range log$\tau$=[0,-5.5], which is the atmosphere from the bottom of the photosphere to the middle chromosphere, up to heights that hydrostatic models estimate to be of about 1000-1500 km above the photosphere. Besides, \citet{Quintero2017} found that the sensitivity  of the \ion{Ca}{2} line to the atmospheric parameters, in particular the photospheric sensitivity to the vector magnetic field, largely increases when  additional photospheric spectral lines are analyzed. 

In their study, \citet[][]{Joshi2018}  analyzed the variation of the atmospheric parameters associated to oscillations manifested as umbral flashes and penumbral waves, but they also estimated the average properties of the vertical gradient of the penumbral magnetic field between the photosphere and chromosphere. They did not describe the fine-scale structure of the observed penumbral field and just reported on a decrease of the magnetic field with a rate of -0.5 G km$^{-1}$ in the vertical direction, in agreement with values in the literature, ranging from -0.3 to -1.0 G km$^{-1}$. 
Earlier studies by \citet{Socas2005a,Socas2005b} of photospheric and chromospheric observations taken 
at the 850 nm spectral range and including \ion{Ca}{2} data, as well,  only showed the complex  topology of sunspot magnetic field with areas of opposite-sign torsion and twist, 
 suggestive of flux ropes of opposite helicity coexisting together in the same spot, with no further details on the fine-scale structure of the field due to the spatial resolution of the analyzed data.

In this paper, we study the  3D magnetic structure of the  penumbra in a large and fully-developed sunspot, 
as derived from simultaneous inversion with the NICOLE code of photospheric Fe I 617.3 nm and chromospheric Ca II
854.2 nm data.  Although the latter data provide an excellent tool for investigating the chromospheric environment \cite[][]{Cauzzi2008}, to date the Ca~II data have been less explored than other  diagnostics to study penumbral regions. 
In this light, the results presented in the following offer new information on the properties of the chromospheric magnetic field in penumbral regions, especially concerning the geometrical heights of different parts of penumbral filaments and the penumbral structure at chromospheric heights above log$\tau$= -2.3, which were not analyzed in previous studies  \citep[][]{Tiwari2017}. 



\section{Data and methods}
\subsection{Observations}
\label{sec:observation and data analysis}

\begin{figure*}
	\centering
	\includegraphics[scale=1.2,clip, trim=0 0 10 13]{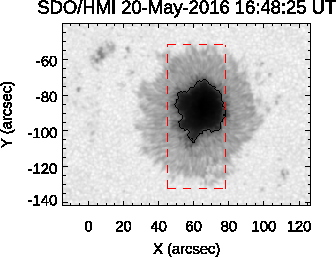}
	\includegraphics[scale=1.2,clip, trim=45 0 10 13]{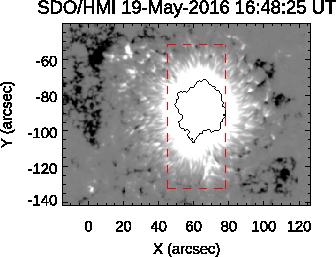}\\
	\caption{Sub-arrays extracted from the continuum filtergram (left panel) and LOS magnetogram (right panel) taken by SDO/HMI on 2016 May 20 at 16:48 UT. Values of the LOS magnetogram are saturated at $\pm$700 G. The red box in both panels indicates the IBIS FoV displayed in Fig. \ref{f2}. Here and in the following figures, solar North is at the top, and West is to the right.}
	\label{f1}
\end{figure*} 

\begin{figure*}
	\centering
	\includegraphics[scale=0.95,clip,trim=0 30 0 0]{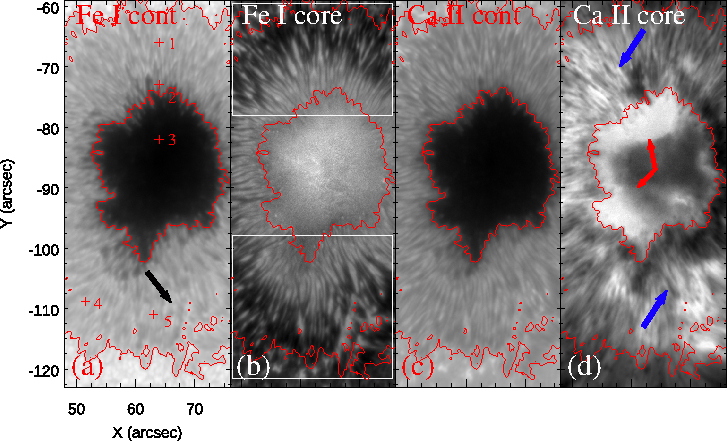}\\
	\includegraphics[scale=0.95,clip, trim=0 0 0 0]{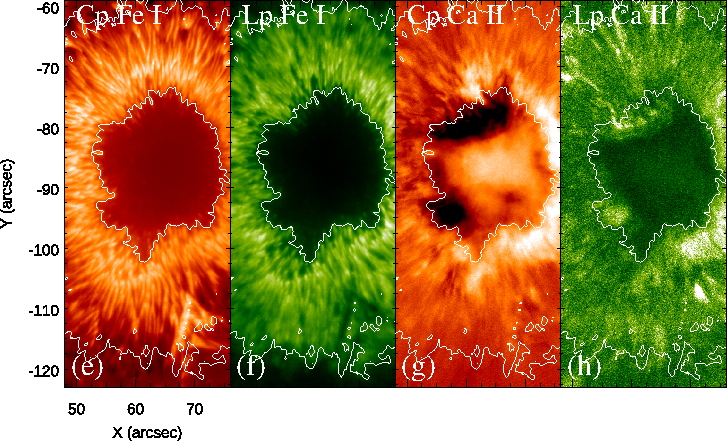}
	\caption{Top panels: Maps of the Stokes I in the \ion{Fe}{1} 617.3 nm a) line-continuum, b) line-core, and in the  \ion{Ca}{2} 854.2 nm c) line-continuum, and d) line-core data. Bottom panels: Maps of the \textit{C$_p$} and \textit{L$_p$} polarization signals  derived from the  Fe~I  and  Ca~II line data. The cross symbols in panel a) indicate the location of the pixels considered in Figure \ref{f3}, while the sub-arrays marked with solid lines in panel b) indicate the regions considered in Figs. \ref{f5} and \ref{f7}. The red contours in all of the panels show the UP boundary and the outer penumbra boundary set at I$_{c}$=0.4 and I$_{c}$=0.9, respectively, where I$_{c}$ is the mean value of Stokes I in a quiet Sun region. The arrows indicate the regions described in the text.}
	\label{f2}
\end{figure*}

We analyzed data acquired on 2016 May 20, with the Interferometric Bidimensional Spectrometer \citep[IBIS,][]{Cav06} at the Dunn Solar Telescope (DST) of the National Solar Observatory. These observations  concern the large and mature sunspot of active region (AR) NOAA 12546 at the time located near disk center at (S07,W07) and $\mu \approx$0.97, where $\mu$ is the cosine of the heliocentric angle. The data consist of full-Stokes measurements taken from 13:53 UT to 18:17 UT along the Fe I 617.3 nm  and Ca II  854.2 nm lines, with a cadence of 48 s. The spectral ranges of the Fe~I and Ca~II measurements span from 617.1 nm to 617.5 nm and from 853.6 nm to 854.8 nm, respectively. The polarimetric data were taken at 21 spectral points in each line, with a spectral sampling of 20 m\AA~and 60 m\AA~for the Fe I 617.3 nm and Ca II 854.2 nm lines, respectively. The field of view (FoV) was 500 $\times$ 1000 pixels with a pixel scale of 0\farcs08. The above measurements were complemented with simultaneous broad-band images taken at $633.32 \pm 5$ nm on the same FoV of, and with the same exposure time of the polarimetric data.
For further details on the analyzed observations see  \citet[][]{Stanga2018}.

The sunspot was observed with the adaptive optic system (AO, Rimmele 2004) of the DST locked and running on the center of the umbral region. The spatial resolution of the data is 0.16\arcsec and 0.23\arcsec ~for the Fe I and Ca II  line measurements, respectively. 

In this paper we show results derived from analysis of the best scan in the long-duration data-set obtained during the observations (318 scans available). According to rms values in quiet Sun regions and to all the measurements taken along the sampled lines, the best scan corresponds to the nr 191 of the series, whose acquisition started at 16:52 UT.

The observations were calibrated  by using standard procedures for flat-fielding, dark subtraction, and polarimetric calibration.
Moreover, 
the data were restored for seeing-induced  degradation left over by the telescope AO system with application of the Multi-Object Multi-Frame Blind Deconvolution \citep[MOMFBD,][]{vannoort2005} technique to the calibrated measurements.
In addition to the seeing degradation, the observations analyzed in our study also suffer from stray-light contamination due to instrument design. According to \citet[][]{ReardonCavallini}, at the Fe~I and Ca~II spectral ranges the level of the IBIS stray-light is $\approx$  2\% and 1.3\%  of the instrumental transmittance, respectively. We account for this degradation of the data in the NICOLE inversion described in the following,  by allowing the macroturbulence parameter to vary in the processing, without considering the instrumental stray-light point-spread-function in the computations. 

AR NOAA 12546  appeared on the East solar limb on 2016 May 13. It consisted of an isolated, large, and fully-developed sunspot  with a leading unipolar magnetic field configuration ($\alpha$-type) of  positive polarity, and a nearby plage region, with trailing negative polarity fields. It is worth nothing that this is a peculiar sunspot due to its dimension and magnetic field strength (exceeding 4 kG, as resulting from Hinode/SOT-SP level 2 data, not shown here). In fact, in a recent work by \citet{Livingston2015} is shown that in the 2010-2015 period there were only few sunspot with an umbral magnetic field exceeding 3500 G.  

Figure \ref{f1} shows sub-arrays extracted from the continuum filtergram and LOS magnetogram taken by the Helioseismic and Magnetic Imager \citep[HMI,][]{Sch12}  onboard the Solar Dynamics Observatory \citep[SDO,][]{Pesnell2012} on 2016 May 20, at 16:48 UT, i.e. close in time to the IBIS observations analyzed in our study. The SDO/HMI observations were taken in the Fe~I  617.3 nm line with a pixel scale of about 0.5\arcsec.

The red box in  Fig. \ref{f1} displays the FoV of the IBIS observations, whose examples are given in Fig.  \ref{f2}. The IBIS FoV, which is centred on the sunspot barycentre, covers the entire umbra and most of the penumbral region of the studied spot, especially in the North and South directions.

Figure \ref{f2} (top panels) shows IBIS intensity images obtained in the continuum and in the core of the Fe~I 617.3 nm and  Ca~II 854.2 nm lines. In order to make the small-scale structures in the FoV clearer, we display both the \ion{Fe}{1} and \ion{Ca}{2} line-core intensity maps with the values normalized to the local continuum intensity.

The IBIS Fe~I line-continuum intensity data (Fig. \ref{f2}a), and more thoroughly the co-temporal SDO/HMI continuum filtergram in Fig. \ref{f1}, show that the studied penumbra is almost circular and rather homogeneous, except for a small sector in the South-West direction (shown by the black arrow in panel a). 
The Fe~I line-core intensity data (Fig. \ref{f2}b) display different average values in the inner and outer parts of the penumbra, with a decrease of the line-core intensity values when moving away from the spot center.  
The \ion{Ca}{2} line-continuum intensity map (panel c) shows minute deviations to the corresponding \ion{Fe}{1} map.
The intensity map obtained in the line-core of the Ca~II data (Fig. \ref{f2}d) shows the super-penumbra \citep{Loughhead68} near the edges of the FoV, especially at the North-West and South-East directions. Unfortunately, 
the IBIS FoV covers only a small part of the super-penumbral region. 
The same data also display two bright regions, located to the North-East and South-West sides of the FoV (shown by the two blue arrows in Fig. \ref{f2}d), the latter corresponding with a region where the distribution of interlaced filaments is less regular than in other penumbral sectors (marked by the black arrow in Fig. \ref{f2}a). 

Figure \ref{f2} (bottom panels) shows the maps of the mean circular $C_{p}$  and linear $L_{p}$  polarization signals derived from the Fe~I and Ca~II line data. 
These maps were computed pixelwise following \citet{Guglielmino2012} by using the  formulae: 
\begin{eqnarray}
C_{p}=\frac{1}{10 \left \langle I_{c} \right \rangle} \sum_{i=5}^{15} \epsilon_{i}  |{V_{i}}|  \\
L_{p}=\frac{1}{10 \left \langle I_{c} \right \rangle} \sum_{i=5}^{15} \sqrt{Q^{2}_{i}+U^{2}_{i}}
\end{eqnarray} 
where $\left \langle I_{c} \right \rangle$ is the average value of Stokes I in the line-continuum in a quiet region, $\epsilon=1$ for the first 5 spectral positions of the line sampling,  $\epsilon=-1$ for the last 5 positions, and $\epsilon=0$ for the line center position, and i runs from the $5^{\mathrm{th}}$  to the $15^{\mathrm{th}}$  wavelength positions.

The $C_{p}$ (Fig. \ref{f2}e) and $L_{p}$ (Fig. \ref{f2}f) maps derived from the Fe~I data clearly show the spine/intra-spine structure of the penumbra observed in the photosphere. However, signatures of the same structure are also found in both the $C_{p}$ (Fig. \ref{f2}g) and $L_{p}$ (Fig. \ref{f2}h) maps obtained from the Ca~II data. 
The intense brightening in the $C_p$ map localized near the umbra-penumbra (UP) boundary (see the two red arrows in panel Fig. \ref{f2}d), defined by the threshold in the Fe I line-continuum intensity I$_{c}$=0.4,  is due to wave processes that will be analysed in a future paper. 


\subsection{Data Inversion}

\begin{figure*}
	\centering
	\includegraphics[scale=0.6,clip, trim=0 0 0 0]{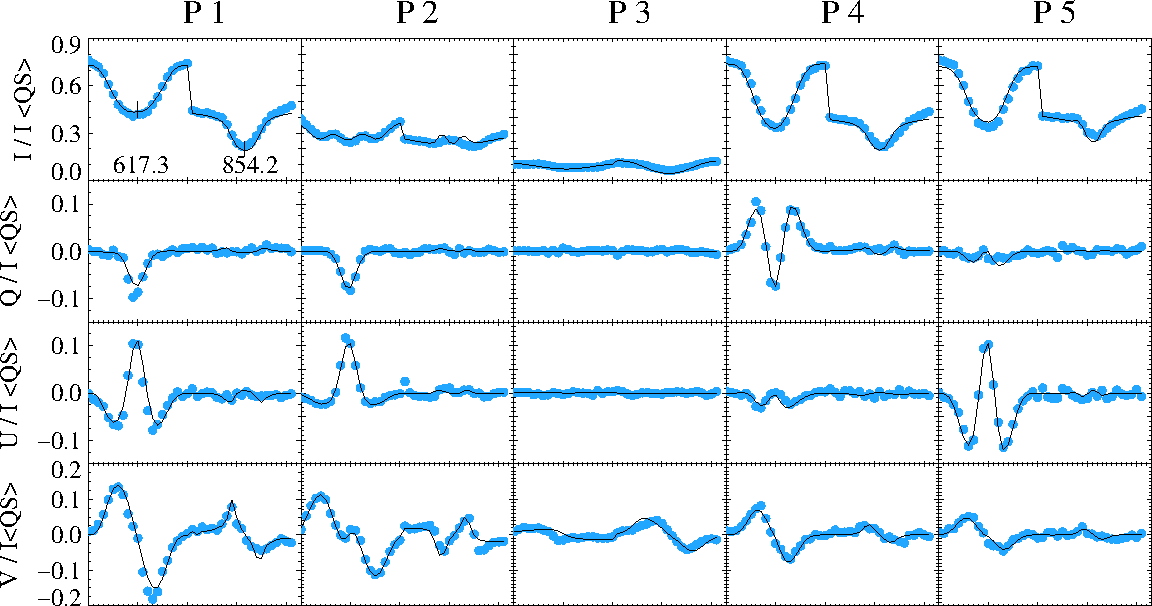}
	\caption{Examples of observed (filled light blue circles) and inverted (black solid line) Stokes profiles for the image pixels belonging to the penumbra (P1, P4, and P5), UP-boundary (P2), and umbra (P3) at the  locations shown  in Fig. \ref{f2} (panel a). }
	\label{f3}
\end{figure*}

We used the non-LTE inversion code NICOLE  \citep[][]{Nicole2015} to derive the physical parameters in the observed region at the atmospheric heights sampled by the analyzed data. NICOLE solves the non-LTE radiative transfer  in a plane-parallel geometry under statistical equilibrium, by assuming complete redistribution in angle and frequencies of scattered photons to compute intensities. 
We refer the reader to the above paper for further details. 


We show in the following the results derived from inversions of the Fe~I and Ca~II  lines simultaneously. We used five equidistant nodes located from  log$\tau$=-7 and log$\tau$=1 for temperature, three nodes for each component of the vector magnetic field (B$_x$, B$_y$, and B$_z$), two nodes for the line-of-sight (LOS) velocity, and one node for both the microturbulence and macroturbulence. This setting of the nodes, which  resulted from test inversions performed to best reproduce the observed spectra, is compliant with that used by \citet[][]{Robustini2018} to derive the magnetic field topology in chromospheric regions hosting jets.  We performed the inversions assuming as the initial guess model a  FALC atmosphere \citep{FAL-C} modified with a constant value of 1.5 kG for B$_z$.

Figure \ref{f3} shows examples of the observed and inverted Stokes profiles for five pixels located at different positions in the FoV, specifically in the umbra,  in the UP boundary, and at three locations in the penumbra. The position of these pixels is marked with crosses in Fig. \ref{f2}a. For all of the considered penumbral pixels, the inverted profiles match  well the observed ones, especially when considering the  Fe~I data that have Stokes Q, U, and V signals with maximum amplitude of about $\approx$ 10\%, 10\%, and 15\%~ the Stokes I values in quiet Sun regions, respectively. The Stokes profiles measured in the  Ca~II line are characterized by lower Stokes Q and U  signals than reported above for the Fe~I data, but the  values measured at the Ca~II are still above the noise level ($\approx$ 5$\times$10$^{-3}$) of the data (typical noise level is 10$^{-3}$; see, e.g. \citet{Lagg2017} and \citet{Stanga2018}). Therefore, the inverted profiles depict reasonably well the observed profiles for all of the analyzed penumbral pixels. On the contrary, for some pixels in the UP boundary and in the  umbra, as those considered in Figs. \ref{f2} and \ref{f3},  the measured Stokes profiles are extremely reduced and even distorted with respect to those reported above. 
For example, the Stokes profiles of the UP boundary pixel considered in Fig. \ref{f3} show  emission line reversals in the blue and red lobes  of both the Fe~I and Ca~II lines. These distorted profiles are likely due to either solar transient events and waves affecting the measurements, or seeing-induced aberrations un-accounted for during the observations and with the data processing. At the UP boundary and in the umbra, the Stokes Q, U, and V signals  are very low. While the Stokes Q and U signals of the chromospheric data in UP pixels are still above the data noise, for some pixels in the umbra, the same signals are close to the noise level, impeding any accurate data inversion. 
Based on these properties of the analyzed data, we focussed our study to the penumbral region and ignored any results derived from the data inversion that concern the transverse and vertical components of the umbral magnetic field. We also assumed results for the transverse  component of the field derived from the data inversion of the UP region with caution. It is worth noting that our data selection is supported by the recent results by \citet{Felipe18}. These authors studied the accuracy of the atmospheric parameters derived from inversion of spectropolarimetric data, including \ion{Ca}{2} observations, obtained with the scanning of spectral lines, as the one analyzed in our study. By using synthetic data from magnetohydrodynamic simulations, \citet{Felipe18} showed that inversion of such data provides an unreliable characterization of the thermodynamic properties of the atmosphere, when the measured Stokes profiles have apparent wavelength shifts and spurious deformations due to propagation of waves in sunspots, as for some of the data analyzed in our study.

Since the studied sunspot was observed close to disk center, the vector magnetic field was considered as returned from the data inversion, i.e. the field values derived from the inversion were not transformed into the solar frame of reference. The $180^{\circ}$ ambiguity in the azimuth direction was also not resolved.

\subsection{Response functions}
\begin{figure*}
	\centering
	\includegraphics[scale=0.36,clip, trim=0 0 0 0]{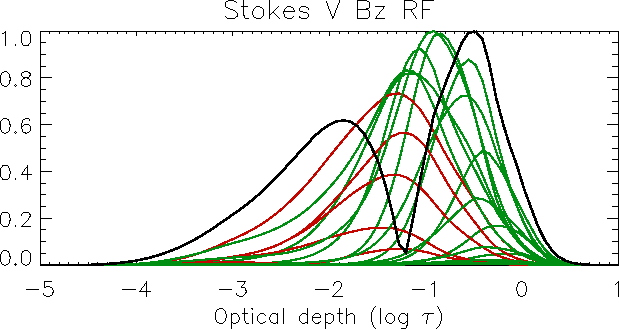}
	\includegraphics[scale=0.36,clip, trim=0 0 0 0]{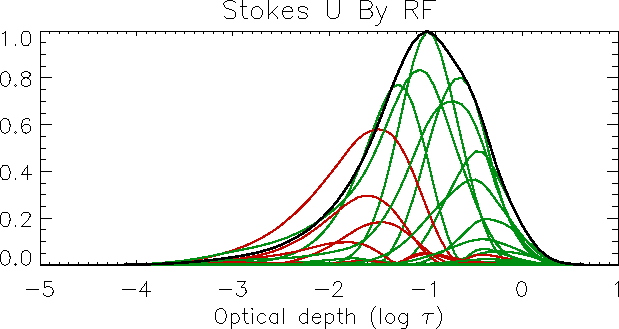}
	\includegraphics[scale=0.36,clip, trim=0 0 0 0]{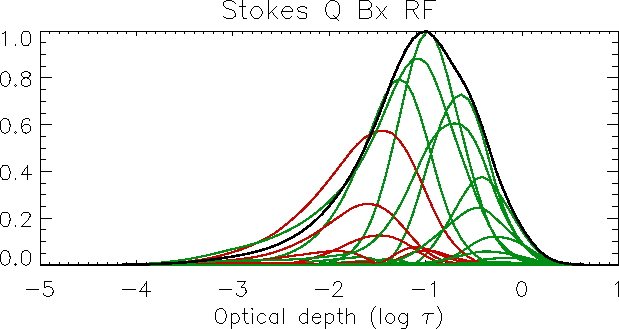}\\
	\includegraphics[scale=0.36,clip, trim=0 0 0 0]{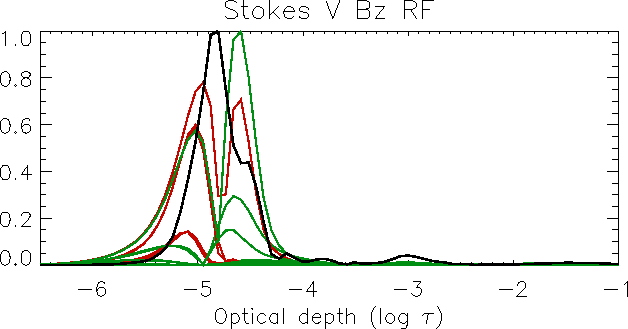}
	\includegraphics[scale=0.36,clip, trim=0 0 0 0]{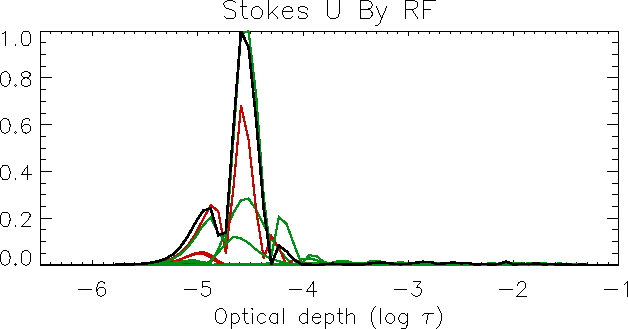}
	\includegraphics[scale=0.36,clip, trim=0 0 0 0]{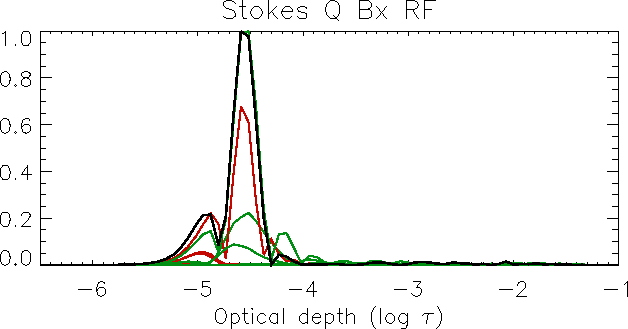}

\caption{RFs for the three components of the vector magnetic field (from left to right: B$_z$, B$_y$, and B$_x$, respectively) obtained perturbing the synthetic model derived from the NICOLE inversion of the Fe~I 617.3 nm (top panels) and Ca~II 854.2 nm (bottom panels) line data, respectively. The different colours designate different spectral regions in the line profiles, specifically the line-wings and line-core are shown with the green and red lines, respectively. The black curves indicate the wavelength integrated RFs. Each RF is normalized to the maximum value of the RF obtained for the given Stokes parameter considered.}
	\label{f4}
\end{figure*}

In order to assess the range of atmospheric heights in which the analyzed data are sensitive to perturbations of the magnetic field, we computed the  response functions (RFs) for the Stokes profiles to perturbations of the atmosphere model derived from the data inversion. 
We followed the method of \citet{Quintero}. 
In particular, we computed $RF^{S}_{B_{i}}$, where S represents Stokes Q, U, and V and $B_{i}$ indicates one of the three magnetic field components, $B_{x}$, $B_{y}$, and $B_{z}$. We computed the RFs for a penumbral area of 10$\times$10 pixels, by averaging the Stokes profiles therein and normalizing the obtained results with respect to the maximum value of the RF obtained for each profile. For a given Stokes parameter, optical depth, and wavelength, RFs values close to unity indicate that the measurements of the corresponding Stokes-parameters are quite responsive to perturbations of the line-forming atmosphere, while low or null values of RFs indicate  that the Stokes measurements are unaffected by  inhomogeneities in the atmospheric parameters, due to changes of the magnetic field. This implies that the data inversion cannot provide reliable information about atmospheric physical quantities in the line-forming regions characterized by low RFs values, since these regions lie outside the sensitivity range of the analyzed data to perturbation of atmospheric parameters. 

Figure \ref{f4} shows plots for some of the estimated RFs that are representative of  all the RFs computed on the atmosphere model returned by the NICOLE inversion. In particular, top panels display 
the RFs for the three components of the magnetic field represented by the Stokes Q, U, and V profiles of the synthetic Fe~I data derived from the inverted atmosphere. These panels show that the analyzed line is sensitive to perturbations of the magnetic field components at atmospheric heights 
ranging from log$\tau$=0.5 to log$\tau$=-4. However, the regions around log$\tau$=0 and those above log$\tau$=-2 show rather low RFs values, suggesting that we cannot retrieve reliable information on the properties of the magnetic field at those atmospheric heights by the analyzed \ion{Fe}{1} data. 
Bottom panels in Fig. \ref{f4} display the RFs derived from analysis of the synthetic Ca~II data. These panels show that the latter data 
are sensitive to variations of the magnetic fields at different atmospheric heights, ranging  from log$\tau$=-1 to log$\tau$=-6. High RF values are obtained for heights between log$\tau$=-4 to log$\tau$=-6, and the maximum of sensitivity is at log$\tau \approx$-4.6.

According to these results, we show in the following estimates of the magnetic field vector at four atmospheric heights corresponding to local maxima of the estimated RFs in the photosphere at log $\tau$=-0.5, -1.0, and -1.5, and in the chromosphere at  log $\tau \approx$-4.6.

\section{Results}\label{sec:result}


\begin{figure*}
	\centering
	\includegraphics[scale=0.5,clip, trim=0 38 0 0]{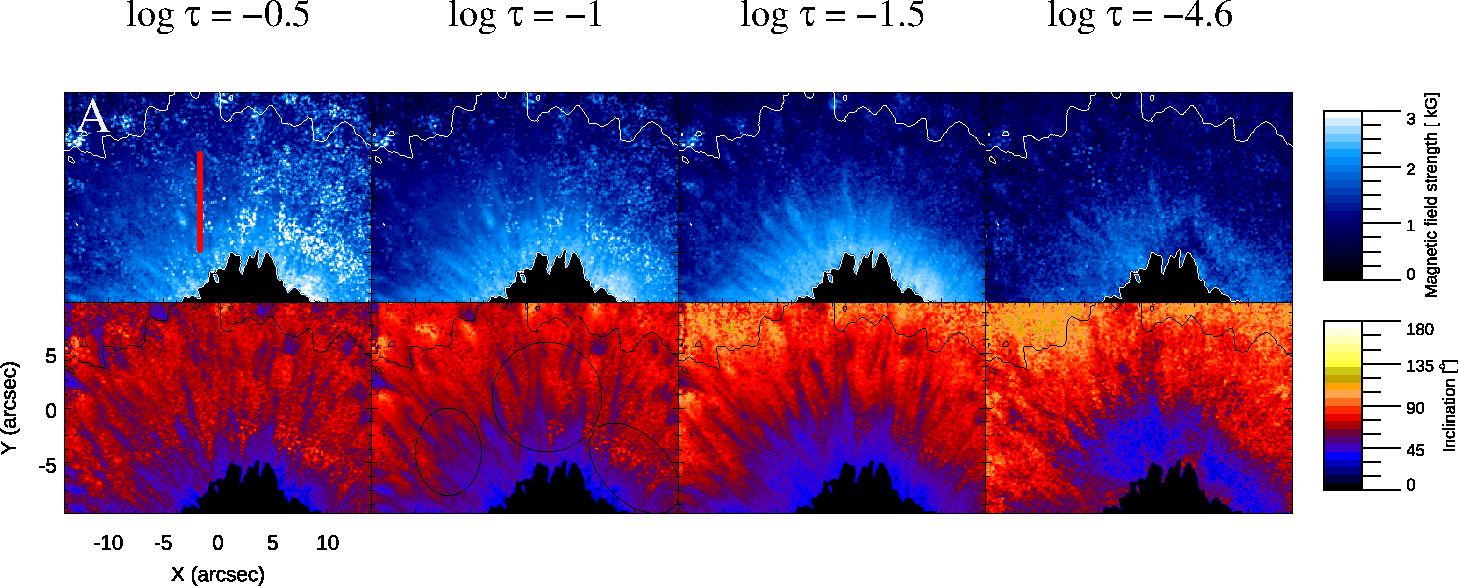}\\
	\includegraphics[scale=0.5,clip, trim=0 0 0 0]{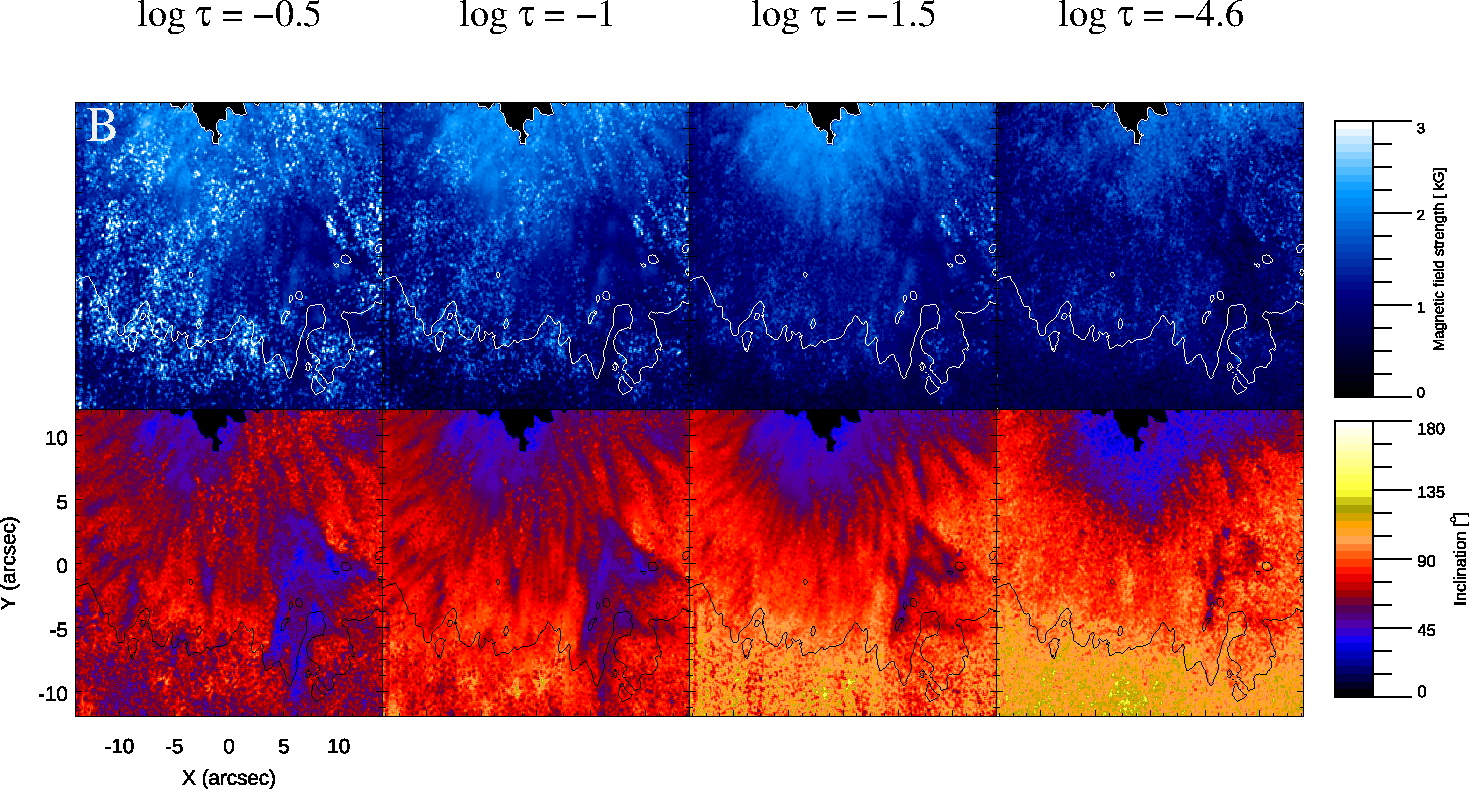}
	\caption{Maps of the magnetic field strength (first and third rows) and  field inclination (second and fourth rows) derived from NICOLE inversion of the \ion{Fe}{1} 617.3 nm and \ion{Ca}{2} 854.2 nm line data. The results shown here refer to the two sub-arrays  displayed in Fig. \ref{f2}b, located in the Northern and Southern sectors of the IBIS FoV.  Black and white contours represent the edge of the umbra and the outer penumbra in the photosphere as defined in Fig. \ref{f2}. The region marked red in the top-left panel is used for the analysis reported in Fig. \ref{f6}. From left to right: The four maps in each row correspond to the results derived from the data inversion at log$\tau$=-0.5, -1., -1.5, and -4.6.  The black ellipses enclose locations where the  small-scale structures described in the main text are seen more clearly. }
	\label{f5}
\end{figure*}

\begin{figure*}
	\centering
	\includegraphics[scale=0.7,clip, trim=0 0 0 0]{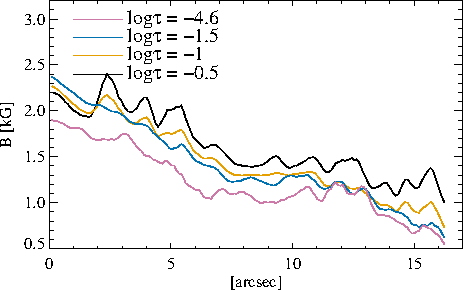}
	\includegraphics[scale=0.7,clip, trim=0 0 0 0]{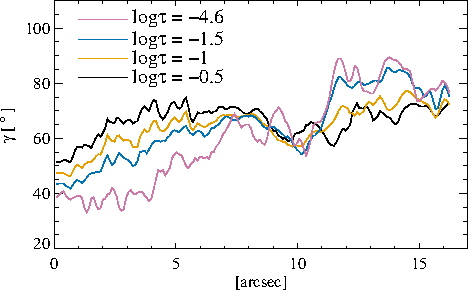}
	\caption{Variation of the magnetic field strength (left panel) and inclination (rigth panel) in the vertical region marked red in Fig.  \ref{f5}. Each profile is obtained by averaging the field values at corresponding distance from the spot barycentre. The distance shown in the x-axis is computed with respect to the UP boundary, moving away from the umbra. Black, orange, blue, and magenta  represent the  field values estimated at log$\tau$=-0.5, -.1, -1.5, and -4.6, respectively. }
	\label{f6}
\end{figure*}

\begin{figure*}
	\centering
	\includegraphics[scale=0.5, clip, trim=0 0 0 0]{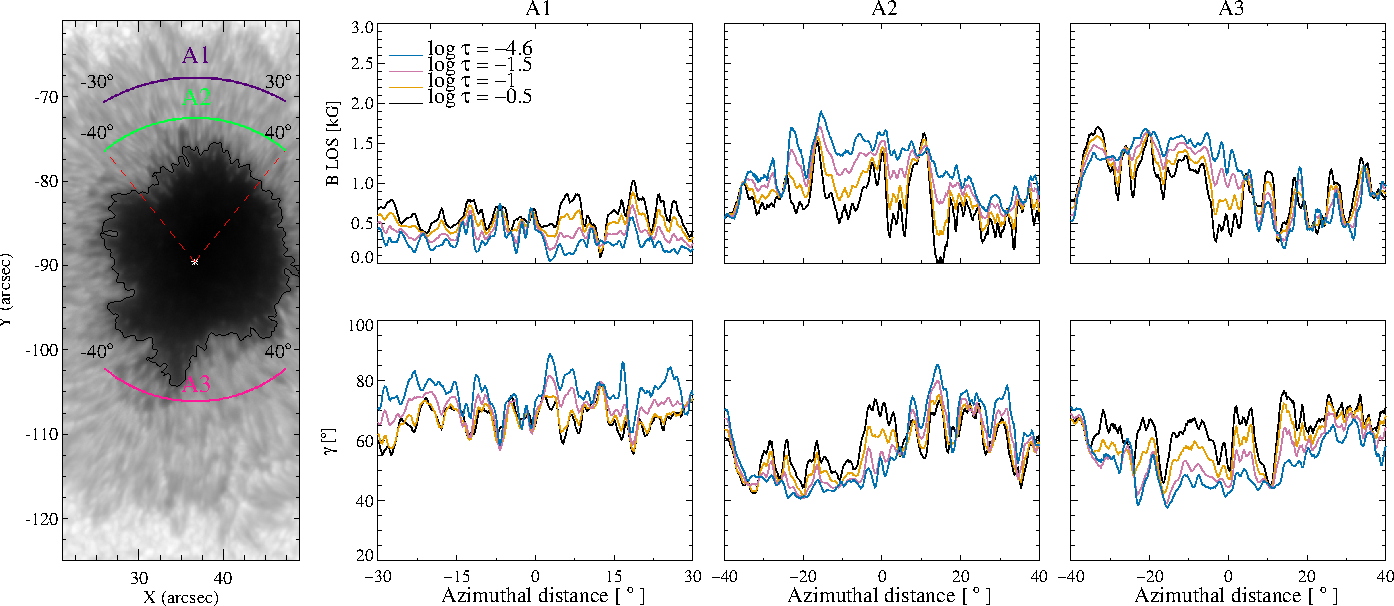}
	\caption{Right-hand panels show the variation of the LOS component (top panels) and inclination (bottom panels) of the magnetic field along the three arcs, A1, A2, and A3 marked with different colours in the Fe~I line-continuum intensity map displayed in the left-hand panel. Results obtained at different values of log$\tau$ are shown with the various colours as displayed in the Legend. Find more details in Sect. 3.}
	\label{f7}
\end{figure*} 


%

\begin{figure*}
	\centering
	\includegraphics[scale=0.7,clip, trim=0 0 0 0]{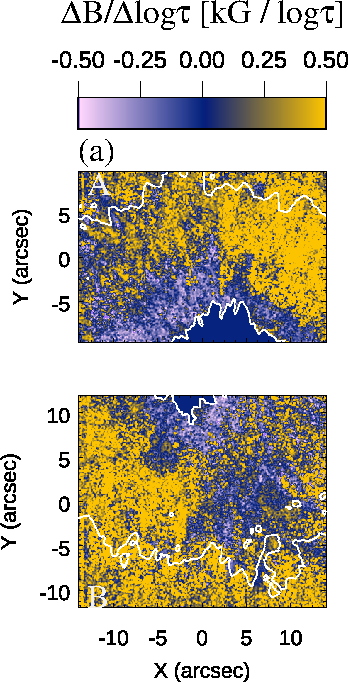}
	\includegraphics[scale=0.7,clip, trim=0 0 0 0]{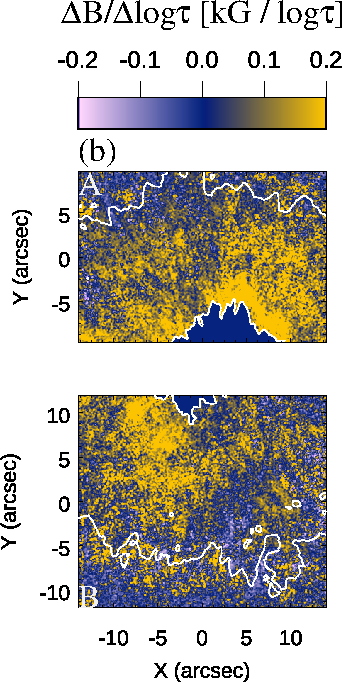}\\
	\includegraphics[scale=0.59,clip, trim=0 0 0 0]{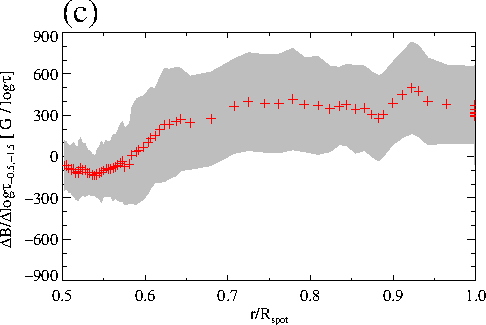}
	\includegraphics[scale=0.59,clip, trim=0 0 0 0]{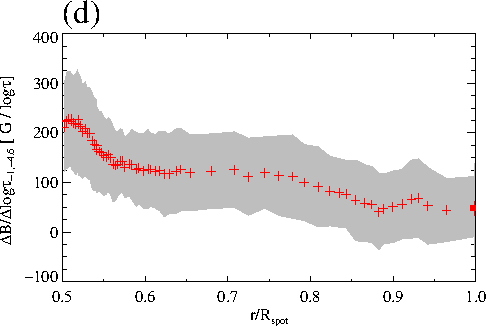}
	\caption{Panels (a) and (b) display the maps of  the vertical gradient of the magnetic field strength by considering two atmospheric heights in the photosphere $(\Delta B / \Delta log \tau)_{-0.5,-1.5}$ (left-hand panels) and two heights representative of the photosphere and chromosphere $(\Delta B / \Delta log \tau)_{-1,-4.6}$ (right-hand panels), for the sub-arrays A (top panels) and B (middle panels) shown in Fig. \ref{f2} (panel b), respectively. White contours indicate the UP boundary and the outer boundary of the penumbra as defined in Fig. \ref{f2}. For the sub-array A, panels (c) and (d) show the variation of the vertical gradient of the magnetic field strength $(\Delta B / \Delta log \tau)_{-0.5,-1.5}$ and $(\Delta B / \Delta log \tau)_{-1,-4.6}$ as a function of $r / R_{spot}$, respectively. The shaded areas represent the standard deviation of the estimated field values.}
	\label{f9}
\end{figure*}



Figure \ref{f5} displays 
sub-arrays 
extracted from the maps of the magnetic field strength and inclination returned from the data inversion.  
These sub-arrays refer to the regions marked with solid lines in Fig. \ref{f2}b,  which are labelled A and B in the following based on their position North and South with respect to the barycentre, of the spot. 

The maps of the magnetic field strength and inclination at log$\tau$=-0.5, -1., -1.5 show the well-known spine/intra-spine structure observed in photospheric penumbrae;  locations where these small-scale features are seen more easily are enclosed by black ellipses overplotted to the inclination map at log$\tau$=-1. Both the studied quantities display rather similar values at all the considered photospheric heights. The maps of the field strength and inclination referring to the chromospheric height log$\tau$=-4.6 show a pattern reminiscent of the small-scale structure observed in the photosphere, but with  
rather different values than those estimated therein. Table \ref{table:1} summarizes the mean and standard deviation of the values of the magnetic field strength and inclination measured inside and outside the penumbra, by pointing out the average variation of these quantities when moving from the photosphere to the chromosphere. For these estimates, the penumbra was identified  as the region where Stokes I of the line-continuum is between 0.4 and 0.9 the average value in quiet Sun regions. It is worth noting that the values of the photospheric magnetic field strength listed in Table \ref{table:1} are comparable with those average magnetic field strength values usually reported for  the umbra of sunpots \citep{Livingston2015}. We stress here that the results presented in the following should be read in light of this.

The maps of the inclination at the photospheric heights log$\tau$=-0.5, -1 display a radially homogeneous distribution of the field with interlaced channel-like features hosting more vertical ad more horizontal fields, respectively. The regular distribution of the inclination fails in the South-West  sector of the penumbra shown in  sub-array B of Fig. \ref{f5}, where the filaments are rather inhomogeneous. 
The maps referring to the chromospheric height at log$\tau$=-4.6  display fields with higher strength in the inner part of the penumbra compared to the fields in the outer part, which however show slight variations along the azimuthal direction. 

Figure \ref{f6} displays the variation of the magnetic field strength (left panel) and inclination (right panel) in the  region marked red in the sub-array A of Fig.  \ref{f5}. This region is  4 pixel wide and 171 pixel long. We computed the variation of  the field strength and inclination along the vertical region moving from the inner to the outer penumbra, by averaging the  values of the magnetic quantities at same distance from the spot barycentre. The distance shown in the x-axis was computed with respect to the UP boundary, moving away from the umbra. Black, orange, blue, and magenta lines represent field values estimated at the atmospheric heights log$\tau$=-0.5, -1., -1.5, and -4.6, respectively. 

When moving from the inner to the outer part of the penumbra, the field strength (left panel of Fig.  \ref{f6}) decreases in the photosphere from 2.3 kG  to 1.5 kG at log$\tau$=-0.5, and from 1.3 kG to 0.7 kG  at  log $\tau$=-1.5; in the chromosphere, it decreases  from  $\approx$ 1.1 kG to 0.6 kG at log$\tau$=-4.6. 
The inner part of the penumbra is characterized by  magnetic fields that become more vertical with the atmospheric height. In particular, close to the UP boundary the values of the field inclination (right panel of Fig.  \ref{f6}) are around 50$^{\circ}$ and  40$^{\circ}$ at log$\tau$=-0.5 and log$\tau$=-4.6, respectively. In contrast, in the outer part of the penumbra the field becomes more horizontal with the atmospheric height. The values of the field inclination measured in the photosphere and chromosphere differ of  about 20$^{\circ}$, reaching   about 90$^{\circ}$ in  the chromosphere.

To further analyze the small-scale spine/intra-spine structure of the penumbra in the photosphere and chromosphere, we considered the properties of the penumbral magnetic field  at different distances from the UP boundary. In particular, we estimated the variation of the LOS component of the field and inclination along the three arcs shown in the Fe~I line-continuum intensity map in Fig. \ref{f7} (left panel). One arc lies in the outher penumbra (hereafter A1), while the other two (hereafter A2 and A3) are located in the inner part of the penumbra}. We consider here the LOS component of the field, instead of the magnetic field strength, because the former quantity shows the spine/intra-spine structure more clearly. We estimated the variation of the field parameters with respect to the azimuthal distance along each arc. 
The filaments in the Northern part of the penumbra sampled by A1 and A2 are  uniformly distributed along azimuthal direction,  while those in the Southern part of the penumbra sampled by A3 are less uniformly arranged, see, e.g., the region centered at [x,y]=[45\arcsec,115\arcsec]. 

Figure \ref{f7} (right panels) displays the LOS component of the magnetic field (top panels) and inclination (bottom panels) along the three studied arcs. The small-scale fluctuation of the field along A1 in the outer penumbra is weak, but still detectable  at all considered atmospheric heights. In the photosphere at log$\tau$=-1 and log$\tau$=-1.5, the LOS component of the field shows interlaced higher and lower values with peak-to-peak changes  of $\approx$300 G; this field fluctuation is reduced to changes of $\approx$150-200 G 
but yet evident at log$\tau$=-4.6. The values of the LOS component of the field along A2 and A3 in the inner penumbra exhibit similar trends as those obtained along A1 in the outer penumbra, but with stronger field values. The field variation along A3 displays a decrease of the average value of the field at the azimuthal distance $\approx$10$^{\circ}$ in the South-West penumbral sector. 

Figure \ref{f7} (bottom panels) shows that, along A2 and A3 in the inner penumbra, the magnetic field inclination decreases passing from the photosphere to the chromosphere. The peak-to-peak variation of the values of the field inclination in these regions decreases from $\approx$ 20$^{\circ}$ at log$\tau$=-0.5 to $\approx$10$^{\circ}$ at log$\tau$=-4.6. The more vertical chromospheric field depicted can be explained with the presence of magnetic canopy at different heights. In detail, when comparing photospheric and chromospheric estimates of the field inclination at the same disc position, we are considering field lines connected to different foot-points in the penumbra \citep{Shahin2017}. The photospheric pattern of the field inclination along A1 in the outer penumbra corresponds well to that observed for the inclination in the middle chromosphere. Besides, along A1 the magnetic field becomes more horizontal with the atmospheric height, by assuming values of  $\approx$55-60$^{\circ}$ in the photosphere and up to 90$^{\circ}$ in the chromosphere.


We then studied the vertical gradient of the magnetic field strength, in order to discuss our results with respect to those in the literature. 
Following \citet{Joshi2017a}, we defined the vertical gradient of the magnetic field strength as:
\begin{eqnarray}
\biggl(\frac{\Delta B}{\Delta log \tau}\biggr)_{a,b}=\frac{(\Delta B)_{a,b}}{(\Delta log \tau)_{a,b}}=\frac{B(b)-B(a)}{b-a}
\end{eqnarray}  
where \textit{a} and \textit{b} indicate the lower and upper log$\tau$ of compared maps, respectively.

Figure \ref{f9} (panels a and b) shows the maps of ($\Delta B / \Delta log \tau$) obtained by considering different atmospheric heights. Fig. \ref{f9} (left panels) displays results by assuming [\textit{a},\textit{b}]=[log$\tau$=-0.5, log$\tau$=-1.5] in the photosphere, while Fig. \ref{f9} (right panels) considering the heights [\textit{a},\textit{b}]=[log$\tau$=-1, log$\tau$=-4.6] in the photosphere and chromosphere, respectively. Both the left- and right-hand panels display results  for the penumbral regions labelled A (top panels) and B (middle panels) in Figs. \ref{f2} and \ref{f5}.
\newline
The inner part of the penumbra exhibits a ring-like structure in the $(\Delta B / \Delta log \tau)_{-0.5,-1.5}$ map, clearly seen in Fig. \ref{f9} (panel a). In that ring  structure the photospheric vertical gradient of the field strength has negative values, i.e. the field strength decreases with the optical depth. Instead, the field gradient is positive in the outer part of the penumbra. 
\newline
For the region labelled A, Fig. \ref{f9} also displays the radial dependence of the vertical gradient of the magnetic field strength in the photosphere (panel c) and when moving from the photosphere to the chromosphere (panel d).
We derived the radial dependence of the magnetic field by considering the field values at the locations of 80 iso-contours in a smoothed map of the Fe~I line-continuum intensity. This map was obtained by applying a boxcar running average values of intensity over 50$\times$50 pixels. We then computed azimuthal averages of the field strength along the iso-contours. We excluded from this analysis the B sub-array because of the inhomogeneous penumbral filaments.

Figure \ref{f9} (panel c) shows the radial dependence of the vertical gradient of the field strength ($\Delta B / \Delta log \tau)_{-0.5,-1.5}$. At r/R$_{spot}$=0.5 in the inner penumbra the field gradient is negative with values of about -100 G/log$\tau$. From r/R$_{spot}$=0.55, still in the inner penumbra, the field gradient starts to sligthly increase. Moving further away from the umbra, at about r/R$_{spot}$=0.6, the value of the field gradient increases to mean values of about 300 G/log$\tau$; from r/R$_{spot}$=0.6 to r/R$_{spot}$=1.0 the field gradient shows small fluctuations around the same mean value

In contrast, the vertical gradient $(\Delta B / \Delta log \tau)_{-1,-4.6}$  in Fig. \ref{f9} (panel d) displays positive values of about 200 G/log$\tau$ at r/R$_{spot}$=0.5. The gradient slowly decreases when moving away from the umbra, assuming always  positive values till r/R$_{spot}$=1.0. 
From r/R$_{spot}$=0.5 to r/R$_{spot}$=0.6 in the inner penumbra, the field gradient has an average value of about 
 170 G/log$\tau$. From r/R$_{spot}$=0.6 onwards the gradient maintains to about  100 G/log$\tau$. Overall, in the whole penumbra the vertical gradient slightly decreases, i.e., the magnetic field increases with optical depth moving out from the sunspot centre.
 
The above values of vertical magnetic field gradient are summarized in table \ref{table:2}.



\begin{table}[t]
\caption{Magnetic field strength and inclination}          
\label{table:1}      
\centering        
\begin{tabular}{c c c c c }            
\hline
\hline
                &                      &                       &                                 &                               \\
log$\tau$            & $\overline{B}_{p}\pm \sigma$   & $\overline{B}^{out}\pm \sigma$  & $\overline{\gamma}_{p}\pm \sigma$         & $\overline{\gamma}^{out}\pm \sigma$ \\
                & [kG]                 & [kG]                  & [$^{\circ}$]                    & [$^{\circ}$]               \\ 
                \hline
                &                      &                       &                                 &                            \\    
                       
-0.5  & 1.7 $\pm$ 0.4        & 1.6   $\pm$ 0.4       &  60  $\pm$ 9\hphantom{5}        & \hphantom{1}63   $\pm$ 9\hphantom{5}         \\    
                &                      &                       &                                 &                            \\   
-1    & 1.6 $\pm$ 0.4        & 0.9   $\pm$ 0.4       &  61  $\pm$ 9\hphantom{5}        & \hphantom{1}78   $\pm$ 9\hphantom{5}             \\
                &                      &                       &                                 &                            \\  
-1.5  & 1.6  $\pm$ 0.4       & 0.8   $\pm$ 0.4       &  64  $\pm$ 9\hphantom{5}        & \hphantom{1}95   $\pm$ 9\hphantom{5}            \\
                &                      &                       &                                 &                            \\
-4.6  & 1.2   $\pm$ 0.5      & 0.8  $\pm$ 0.5        &  68  $\pm$ 15                   &  104  $\pm$ 15           \\
                &                      &                       &                                 &                            \\
\hline 
\end{tabular}
\begin{tablenotes}
      \footnotesize \item[1] $\overline{B}_{p}\pm \sigma$: Mean $\pm$ standard deviation of the values of the magnetic field strength  in the penumbra; \item[2] $\overline{B}^{out}\pm \sigma$: Mean $\pm$ standard deviation of the values of the magnetic field strength  outside the penumbra; \item[3] $\overline{\gamma}_{p}\pm\sigma$: Mean $\pm$ standard deviation of the values of the magnetic field inclination  in the penumbra; \item[4] $\overline{\gamma}^{out}\pm\sigma$: Mean $\pm$ standard deviation of the values of the magnetic field inclination outside the penumbra 
    \end{tablenotes}
\end{table}

\begin{table}[t]
\caption{Vertical gradient of the magnetic field strength}          
\label{table:2}      
\centering        
\begin{tabular}{c l l}            
\hline
\hline
                          &                                                                &                                                        \\
    r/R$_{spot}$          & $\biggl(\frac{\Delta B}{\Delta log \tau}\biggr)_{-0.5,-1.5}$   & $\biggl(\frac{\Delta B}{\Delta log \tau}\biggr)_{-1,-4.6}$ \\
    		      	  & G /log$\tau$						   & G /log$\tau$\\
\hline                      
		          &                                                                &                                 \\
$\leq 0.6$                &     -110                                       		   & 170   \\ 
                          &                                                                &                                               \\            
$> 0.6$ and $\leq 0.9$  &  \hphantom{-}300                                               & \hphantom{1}97                    \\    
                          &                                                                &                                                     \\
$> 0.9$ and $\leq 1.0$  &  \hphantom{-}350                                               & \hphantom{1}46                     \\                                           
			  &                                                                &                                      \\
\hline 

\end{tabular}
\begin{tablenotes}
      \footnotesize \item[1] $(\Delta B / \Delta log \tau)_{-0.5,-1.5}$: Mean value of the photopheric vertical gradient of the magnetic field strength between log$\tau$ = -0.5 and log$\tau$ = -1.5.
     \item[2] $(\Delta B / \Delta log \tau)_{-1,-4.6}$: Mean value of the vertical gradient of the magnetic field strength between log$\tau$ = -1.0 and log$\tau$ = -4.6. 
    \end{tablenotes}
\end{table}

\section{Discussion}

We analyzed the 3D structure of the magnetic field in the penumbra of a large and fully-developed  sunspot, as inferred from inversions of spectro-polarimetric data taken along the \ion{Fe}{1} 617.3 nm and \ion{Ca}{2} 854.2 nm lines.

\begin{figure*}
	\centering
	\includegraphics[scale=0.70,clip, trim=0 0 0 0]{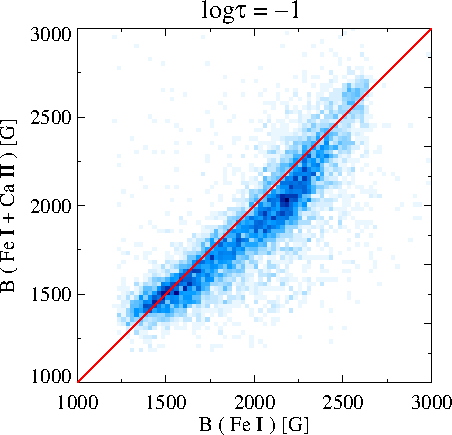}
	\includegraphics[scale=0.70,clip, trim=0 0 0 0]{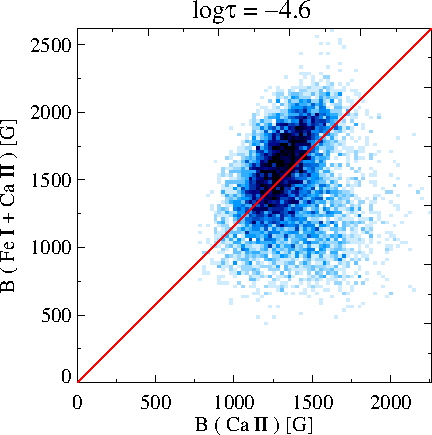}
	\caption{Comparison of the values of the magnetic field strength derived through the inversions of the Fe~I and Ca~II data simultaneously and of either the Fe~I (left panel) or Ca~II (right panel) data, at log$\tau=-1$ and log$\tau=-4.6$, respectively.  Red line in both panels indicates one-to-one correspondence. }
	\label{f10}
\end{figure*}

In order to verify the accuracy of our inversion results, we  applied the NICOLE code to the Fe~I and Ca~II data separately, as recently done by other authors, e.g., \citet{Joshi2018}, and studied the results obtained with respect to those derived from the data inversion of both lines simultaneously. In this computational test we assumed three nodes located from  log$\tau$=-7 and log$\tau$=1 equidistantly for temperature, one node for each component of the vector magnetic field (B$_x$, B$_y$, and B$_z$), one node for the LOS velocity, and one node for both the microturbulence and macroturbulence, respectively. This setting of the data inversion is similar to that employed by \citet{Joshi2018}. Figure \ref{f10} shows the comparison of magnetic field strength values retrieved through the inversion of both the Fe~I and Ca~II lines data simultaneously and of the the data of each line separately. This comparison is displayed at the optical depths where the RFs estimated as in Sect. 2.3 get  maximum values, corresponding to 
log$\tau$=-1 and log$\tau$=-4.6  for the Fe~I and Ca~II measurements, respectively.
The red line in the panels represents one-to-one correspondence.

Figure \ref{f10} (left panel) displays that at log$\tau$=-1 the values of the  magnetic field strength retrieved through the inversion of both lines simultaneously are in a close one-to-one correspondence to the values of the field obtained through the inversion of only the Fe~I data. The relation between the compared series displays that fields whose strength ranges from 1.6 kG to 2.1 kG have slightly lower values ($<$2.5\%) estimated  by the simultaneous inversion of both lines with respect to findings from inversion of only Fe~I data. At log$\tau$=-4.6 (Fig. \ref{f10}, right panel) the correspondence between the values of the  magnetic field strength retrieved through the inversion of both lines simultaneously are in reasonable agreement with those derived from the inversion of only the Ca~II data, but the correspondence of values is less linear in comparison with the previous case. Moreover, for most of the image pixels, the simultaneous inversion of the data of both lines seems to produce slightly higher estimates of the field strength than derived from inversion of the Ca~II data only. However, the difference of results derived from the two inversion methods can be as high as 40\% for the pixels characterized by large values of field strength. Besides, there are image pixels that exhibit lower field strengths (up to 30\%) with respect to those obtained from inversion of the Ca~II line alone. 

Based on  these results, we consider that the stratification of the magnetic field
obtained from our data inversion is reliable and accurate enough to discuss the findings from our study with respect to those in the literature. Furthermore, our findings derive from analysis of photospheric line data in addition to chromospheric \ion{Ca}{2} line measurements. According to  \citet{Quintero2017} the method employed in our study can greatly enhance the sensitivity of the analyzed data to the atmospheric parameters at lower heights.

We found that the  magnetic field strength and inclination in the penumbra display a small-scale spine/intra-spine structure at all of the atmospheric heights considered, from the photosphere to the middle chromosphere, though in the latter region the small-scale pattern is attenuated with respect to the one observed deeper in the atmosphere.

The  channel-like structure of the vector magnetic field also weakens when moving along filaments away from the umbra. At the studied chromospheric height we found signatures of the tails of penumbral filaments; in particular, at  the  outer border of the penumbra, we found patches of the horizontal magnetic fields with inclination of $\approx$ 90$^{\circ}$, and of field patches falling down to the photosphere with inclination of $\approx$ 105$^{\circ}$.

Observations of the fine structure of penumbra in the chromosphere were recently presented by \citet{Joshi2016} on the base of data acquired at the spectral region of the  \ion{He}{1} triplet  with the 1.5 m-class GREGOR telescope \citep[][]{Schm12}. Those authors found an azimuthal variation of the magnetic field inclination in the upper chromosphere, co-spatial with the spine/intra-spine pattern of inclination seen in the photosphere. The reported peak-to-peak variations of the inclination is about $\approx$10$^{\circ}$-15$^{\circ}$ in the chromosphere, compared to the peak-to-peak variation of $\approx$20$^{\circ}$-25$^{\circ}$ in the photosphere. 
\citet{Joshi2016} did not found any azimuthal  variation of the magnetic field strength in the chromosphere, attributing this result  to the higher magnetic pressure of plasma with respect to the gas pressure at atmospheric heights sampled by the \ion{He}{1} triplet.  

In contrast, the results derived from our study reveal a small-scale spine/intra-spine structure of both the magnetic field strength and inclination in the middle chromosphere at the atmospheric heights sampled by the analyzed \ion{Ca}{2} 854.2 nm data. Neverthless, the small-scale field pattern is less structured when moving from the photosphere to the chromosphere; the fine-structuring of the field observed in the chromosphere coincides spatially well with that in the photosphere. We found  typical peak-to-peak variations of the field strength and inclination of $\approx$300 G and $\approx$20$^{\circ}$, respectively, in the photosphere at log$\tau$=-0.5, and of  $\approx$ 200 G and $\approx$ 10$^{\circ}$ in the chromosphere at log$\tau$=-4.6. The above values of the peak-to-peak variation of the magnetic field inclination are in agreement with those reported by \citet{Joshi2016}. 

We also computed the photospheric vertical gradient $(\Delta B/ \Delta log \tau)$ $_{-0.5 , -1.5}$ in the Northern and Southern sectors of the studied penumbra. The maps of this quantity show a ring like structure in the inner penumbra (from r/R$_{spot}$=0.5 to r/R$_{spot}\approx$0.6) similar to the one reported by \citet[][]{Joshi2017a} from analysis of photospheric data taken with the VTT/TIP-2 (between log$\tau$=-2.3 and log$\tau$=0.0) and with the Hinode/SOT-SP (between log$\tau$=-0.9 and log$\tau$=0.0). However, our average photopheric $(\Delta B/ \Delta log \tau)$ is higher than that reported by \citet[][]{Joshi2017a}. In particular, we found  values of the field gradient of about 300 G/log$\tau$ when considering the  photospheric heights log$\tau$=-0.5, -1.5, while \citet[][]{Joshi2017a} reported a value of about 50 G/log$\tau$ considering data taken with the VTT/TIP-2 from log$\tau$=0.0 and log$\tau$=-2.3, and values of about 150 G/log$\tau$ and 100 G/log$\tau$ considering data taken with the Hinode/SOT-SP from log$\tau$=0.0 and log$\tau$=-0.9 and from log$\tau$=-0.9 and log$\tau$=-2.5, respectively. It is also worth nothing that our tests on field parameters estimated by the inversion of the data suggest that the field strength values derived from our study are, on average, smaller than those derived from inversion of photospheric data as in the previous studies. In addition, we computed  the vertical gradient between the chromosphere  and photosphere  at the atmospheric heights where the observations analyzed in our study have maximum sensitivity to magnetic field variations. We found  that the magnetic field strength decreases with optical depth in the penumbra. In particular, we report values of the field gradient of about 100 G/log$\tau$ when comparing the field strength at photospheric and chromospheric heights log$\tau$=-1, -4.6. 
By assuming the difference in the formation height of the \ion{Fe}{1} 617.3 nm and the \ion{Ca}{2} 854.2 nm lines derived from analysis of  
the  phase difference in the propagation of the \textit{p}-modes in the region studied by \citet[][]{Stanga2018}, this difference being of $\approx$ 300 km if the magneto-acoustic velocity is 7 km/s, the latter value  of the vertical gradients of the magnetic field strength of 100 G/log$\tau$  corresponds to a gradient of $\approx$ 0.3 G km$^{-1}$. 
 This result lies in the range of values  reported by \citet{Joshi2017b}, and much earlier by \citet{Ruedi1995}, from analysis of chromospheric data taken in the spectral region of the  \ion{He}{1} 1083 nm triplet.



\section{Conclusions}

We studied the  magnetic field strength and inclination in a sunspot penumbra, analyzing spectro-polarimetric measurements at the \ion{Fe}{1} 617.3 nm and \ion{Ca}{2} 854.2 nm lines. The analyzed data sample the solar atmosphere from the deep photosphere to the middle chromosphere. In order to infer the physical properties of the penumbral plasma at different atmospheric heights, we inverted the available observations with the NICOLE code by processing both line measurements simultaneously. The results presented above derive from  analysis of penumbral data acquired at spectral ranges unexplored in previous studies, and inverted with different methods than those presented in the literature for similar investigations. 
In our maps of  the magnetic field strength and inclination the well known spine/intra-spine structure is clearly seen at all of the atmospheric heights considered. 
Besides, we analyzed the  vertical gradient of the magnetic field strength within different heights in the photosphere, as well as between heights in the photosphere and chromosphere, and found larger values for this quantity than reported from earlier studies. We draw attention on that, based on the values of the magnetic field strength estimated in the umbra, the gradient derived from our study may  represent  
a peculiar case due to the extreme magnetic pressure reached in the umbra.

We presented results derived from inversion of photospheric and chromospheric observations, limiting the analysis to the observed regions where measured and inverted profiles are in good agreement. This is not the case for, e.g., some observed umbral and UP boundary regions, which however were not the focus of the present work.   
Our study offers new observational constraints on the 3D magnetic structure of penumbral regions from analysis of data that sample chromospheric heights not considered in previous studies. However, it also 
%
clearly manifests the need for more data, as well as for simultaneous observations of penumbral regions by using  multiple spectral diagnostics of the photosphere and chromosphere, in order to fully depict the 3D nature of the magnetic field at higher atmospheric heights. Indeed, these observing capabilities are foreseen to be available soon at the Swedish Solar Telescope, currently one of the most  highly resolving solar telescopes, with the realization of the HeSp spectrometer working at the spectral range of the \ion{He}{1}, in addition to the already operative CRISP/CHROMIS spectro-polarimeters working in the NUV-Vis bands \citep[][]{Scharmer2017}. These improved observing capabilities are also expected for the next generation 4-m class solar telescopes, as the Daniel K. Inouye Solar Telescope \citep[DKIST,][]{Kei10} and European Solar Telescope \citep[EST,][]{Col10} under construction and design phase, respectively.


\acknowledgments
The authors thank Doug Gilliam (NSO) and Michiel Van Noort (MPS) for their valuable support during the acquisition and MOMFBD processing of the data. S.L.G. and F.Z. thank support by Universit\`a degli Studi di Catania (Linea di intervento 1 e Linea di intervento 2). 
The research leading to these results has received funding from the European Research Council under the European Union's Horizon 2020 Framework Programme for Research and Innovation, grant agreements H2020 PRE-EST (no. 739500) and H2020 SOLARNET (no. 824135). This work was also supported by INAF Istituto Nazionale di Astrofisica (PRIN-INAF-2014). 
SJ acknowledges support from the European Research Council (ERC) under the European Union’s Horizon 2020 research and innovation program (grant agreement No. 682462) and from the Research Council of Norway through its Centres of Excellence scheme, project number 262622.

\end{document}